\documentclass[fleqn,twoside]{article}
\usepackage{espcrc2}

\usepackage{graphicx}

\newcommand{\AmS}{{\protect\the\textfont2
  A\kern-.1667em\lower.5ex\hbox{M}\kern-.125emS}}

\def\bsbar{${\overline{B}_s^0}$}
\def\bs{${B_s^0}$}

\def\bsdec{${B_s^0 \rightarrow J/\psi \phi}$}
\def\GeV  {\ensuremath{\mathrm{ Ge\kern -0.1em V } }}
\def\GeVc2{\ensuremath{\mathrm{ Ge\kern -0.1em V }\kern -0.2em /c^2 }}
\newcommand{\MT}{\ensuremath{M_{\mathrm{top}}}}

\title{Physics at the Tevatron Run II}

\author{D. Zieminska\address  {Indiana University, Physics Department 
        Bloomington, IN 47405, U.S.A}}

\begin{document}

\begin{abstract}
We present recent physics results from the Tevatron Collider experiments CDF and D\O.
\vspace{1pc}
\end{abstract}

\maketitle

\section{INTRODUCTION}
The Tevatron Collider Run II started in March 2002 and is expected
to continue until the end of this decade. The Tevatron and the 
two detectors, CDF and D\O, have been performing  well in 2004,
each experiment is collecting data at the rate 
of $\approx$10 pb$^{-1}$ per week.
The total  luminosity accumulated by August 2004 is $\approx$500 pb$^{-1}$
per detector.
The rich physics program includes the
production and precision measurement of properties of  standard model (SM)
objects, as well as searches for phenomena beyond standard model.
In this brief review we focus on areas of most interest 
to the lattice community. We present
new results on the top quark mass
and their implication for the mass of the SM Higgs boson, 
on searches for the SM Higgs boson, on evidence for the $X(3872)$ state, 
on searches for pentaquarks, and on $b$ hadron properties.
All Run II results presented here are preliminary. 

\section{TOP QUARK MASS}

The experiments CDF and D\O\ published several direct  measurements of
the top quark pole mass, $\MT$, 
based on Run I data (1992-1996).
The ``lepton $+$ jets'' channel yields the most precise determination of
$\MT$. Recently, the
D\O\ collaboration published a new measurement~\cite{Mtop1-D0-l+j-new},
based on a powerful analysis technique yielding  greatly improved precision.
The differential probability 
that the measured variables in any event correspond to the signal
is calculated as a function of $\MT$. 
The maximum in the product of the individual event probabilities 
provides the best estimate of $\MT$.
The critical differences from previous analyses 
in the lepton $+$ jets decay channel lie in 
the assignment of more 
weight to events that are well measured or more likely to correspond to  
$t \bar t$ signal, 
and  the handling of the combinations of final-state objects
(lepton, jets, and imbalance in transverse momentum) 
and their identification with
top-quark decay products in an event. 
The new combined value for the top-quark mass from Run I is 
$\MT  =  178.0\pm4.3~\GeVc2$.

\begin{figure}[htb]
\vspace*{-5mm}
\includegraphics[height=5.8cm,width=8.1cm]  {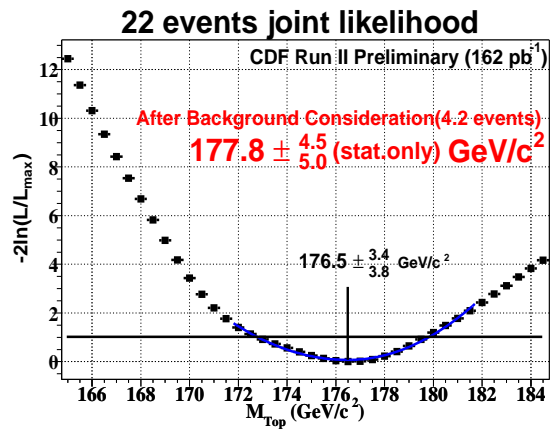}
\vspace*{-1.2cm}
\caption{The joint likelihood of top candidates(CDF).}
\label{fig:cdf_tml}
\end{figure}

In Run II, both collaborations  have been exploring several different techniques 
for $\MT$
measurements. The best single CDF result comes from a dynamic likelihood method
(DLM). The method is similar to
the technique used in Ref.~\cite{Mtop1-D0-l+j-new}.
The result is $\MT = 177.8^{+4.5}_{-5.0} (stat) \pm  6.2 (syst) ~\GeVc2$.
The joint likelihood of the selected events is shown in Fig. ~\ref{fig:cdf_tml}. 
The Run II goal is a 1\% uncertainty on $\MT$. 


\section{SEARCH FOR SM HIGGS BOSON}

The constraints on the SM Higgs ($H$)  boson  mass from
published  measurements, updated to include the new D\O\ top mass
measurement~\cite{Mtop1-D0-l+j-new}, are
$M_H = 117 ^{+67}_{-45}~\GeVc2$, $M_H < 251~\GeVc2$ at 95\% C.L.
The  new most likely  value of $M_H$
is above the experimentally excluded range,
and sufficiently low for $H$ to be observed at the Tevatron.

\begin{figure}[htb]
\vspace*{-5mm}
\includegraphics[height=7.5cm,width=7.8cm]  {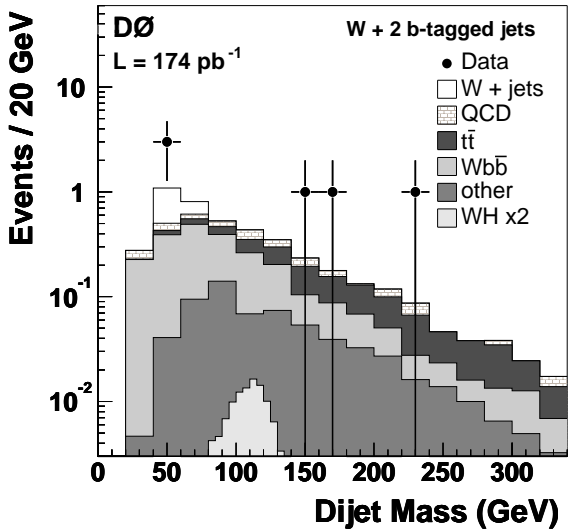}
\vspace*{-1.1cm}
\caption{Distribution of the dijet
invariant mass for $W+2 b$-tagged jets  events,
compared to the expectation (D\O). 
}
\label{fig:d0_wbb_2tag}
\end{figure}

D\O\  has conducted a search for $H$ at $M_H < 140~\GeVc2$ 
in the production channel  
$p \bar{p} \rightarrow WH \rightarrow  e \nu b \bar{b}$. 
The experimental signature of  $WH \rightarrow e \nu b \bar{b}$
is a final state with 
one high $p_T$ electron, two  $b$ jets, and
large missing transverse energy  resulting from
the undetected neutrino.
The dominant backgrounds to $WH$ production
are  $W b \bar{b}$, $t \bar{t}$ and single-top production.
The distribution 
of the dijet mass for events with two $b$-tagged jets is shown in
Fig.~\ref{fig:d0_wbb_2tag}. 
Also shown is the  expected contribution ($0.06$ events)  
from the $b \bar{b}$ decay of a
SM Higgs boson with $M_H =$ 115 $\GeVc2$.
No events are observed in the  dijet mass window of 85--135  $\GeVc2$.
D\O\ sets a limit on the cross section
for $\sigma( p\bar{p} \rightarrow WH) \times B(H \rightarrow b \bar{b}) $
of 9.0 pb at the 95\% C.L.,  for a 115  $\GeVc2$ Higgs boson.
The results for mass points 105, 125, and 135 $\GeVc2$
 are 11.0, 9.1 and 12.2 pb,  respectively.

\begin{figure}[htb]
\vspace*{-1.2cm}
\includegraphics[height=0.33\textheight,width=8.0cm]{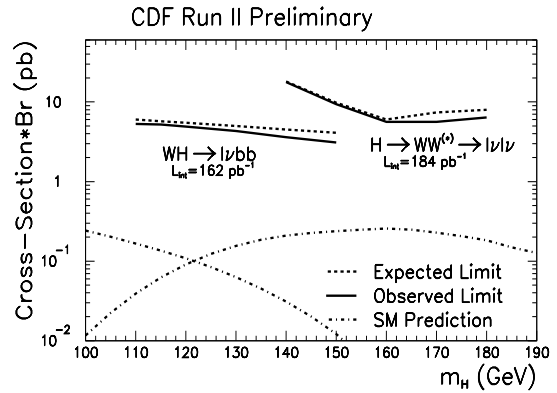}

\vspace*{-1.2cm}
\caption{95\% limits on the $H$ production (CDF).}
\label{fig:cdf_whww}
\end{figure}

CDF  has done  a similar search, allowing either an  electron or a muon  
in the final state.  Both groups have also searched for $H$ produced in
gluon-gluon fusion, with subsequent decay to a pair of $W$ bosons.
The CDF results for both channels  are shown in Fig.~\ref{fig:cdf_whww}.

\section{THE STATE X(3872)}

\begin{figure}[htb]

\includegraphics[height=8.0cm,width=7.5cm]  {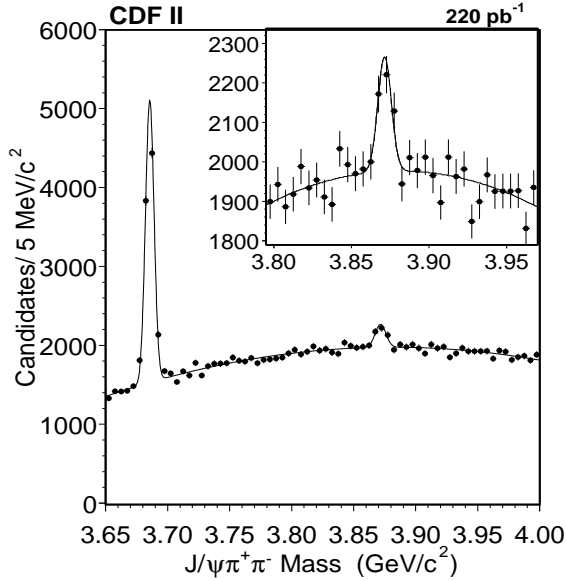}
\vspace*{-1cm}
\caption{The $X(3872)$ signal (CDF).}
\label{fig:cdf_x}
\end{figure}

 The existence of the $X(3872)$ state discovered by 
the Belle Collaboration~\cite{Belle-X}
 has been confirmed 
 in $p \bar{p}$ collisions by  CDF~\cite{cdf-X} (see Fig.~\ref{fig:cdf_x})
and D\O~\cite{d0-X}.
 It is still unclear whether this particle is a $c\bar{c}$ state,
 or a more complex object.  When the data are separated according to
production and decay variables, D\O\  finds no significant
differences between the $X(3872)$ and
the $c \bar{c}$ state $\psi(2S)$.
CDF has analysed the ``lifetime'' distribution of the $X(3872)$ events in order to
quantify what fraction of this state arises from decay of $B$ hadrons, as opposed to
those produced promptly. The authors find that for the selected samples
28.3$\pm$1.0$(stat)\pm$0.7$(syst)$\% of $\psi(2S)$ candidates are from $b$ decays,
whereas 16.1$\pm$4.9$(stat)\pm$2.0$(syst)$\% of $X$ mesons arise from such decays.

%


\section{SEARCH FOR PENTAQUARKS}

\begin{figure}[htb]

\includegraphics[height=0.27\textheight,width=7.6cm]  {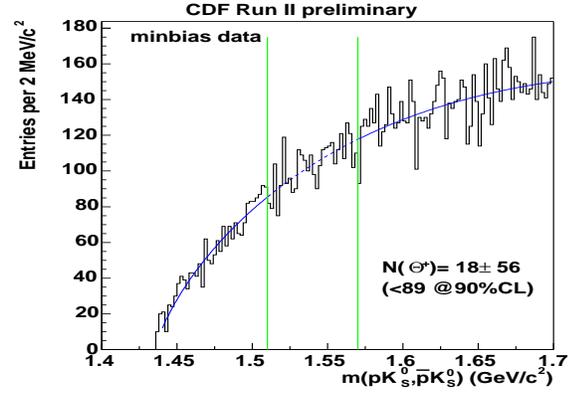}
\vspace*{-1.2cm}

\caption{Invariant mass distribution of an identified proton and a $K^0_s$ candidate. (CDF)
}
\label{fig:pqtheta}
\end{figure}

\begin{figure}[htb]

\vspace*{-0.9cm}
\includegraphics[height=0.25\textheight,width=8.0cm]  {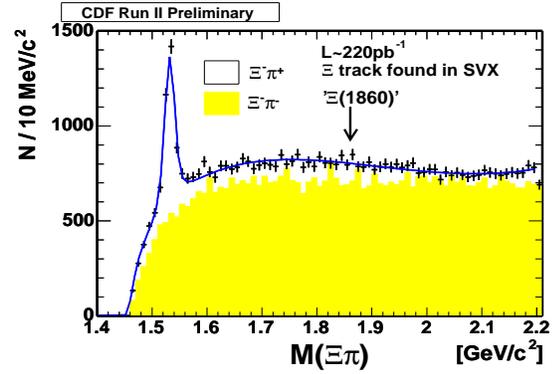}
\vspace*{-1.2cm}
\caption{Invariant mass distribution of the $(\Xi^-,\pi^+)$ system. (CDF) 
}
\label{fig:pqxi}
\end{figure}

\begin{figure}[htb]
\vspace*{-0.9cm}

\includegraphics[height=0.25\textheight,width=7.6cm]  {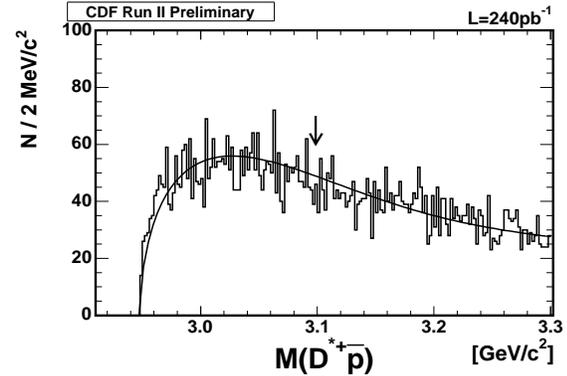}
\vspace*{-1.2cm}
\caption{Mass of the ($D^{*+}\bar p$) system. The arrow indicates the position of 
the $\Theta_c$ state (CDF).}
\label{fig:pqthetac}
\end{figure}

Following reports of evidence for exotic
baryons containing five quarks (pentaquarks), CDF has analysed 
its data for evidence of the following pentaquarks:
$\Theta^+$ ($uud\bar d \bar s$), doubly strange states 
$\Xi_{3/2}$, charmed states $\Theta_c$, and, most recently, 
a state $(udus\bar b)$, dubbed $R^+_s$, through its weak decay to $(J/\psi, p)$. 
With its excellent particle indentification and mass resolution,
CDF has a unique capability to search for  pentaquark states.
The signals of known states: $\phi$, $\Lambda$,
$\Lambda(1520)$, $K^*$, $\Xi$, 
compare favorably with those provided
by the authors of  the pentaquark evidence.
The group finds no evidence for pentaquark states, see Figs 
~\ref{fig:pqtheta},{\ref{fig:pqxi},\ref{fig:pqthetac}.
This can be interpreted as an indication that the pentaquark production 
in $p \bar p$ collisions is heavily suppressed compared to the conventional
hadron production, or as an evidence against the existence of pentaquarks.

\clearpage

\section{RECENT B PHYSICS RESULTS}

\subsection{Spectroscopy}

CDF has measured the mass of $b$ hadrons in exclusive $J/\psi$ channels.
The measurements of the $B_s$ and $\Lambda_b$ (Fig. \ref{fig:masslb})
masses are the current world's best.\\

$m(B^+)$ = 5279.10$\pm$0.41$(stat)\pm$0.36$(syst)$,

$m(B^0)$ = 5279.63$\pm$0.53$(stat)\pm$0.33$(syst)$,

$m(B_s)$ = 5366.01$\pm$0.73$(stat)\pm$0.33$(syst)$,

$m(\Lambda_b)$ = 5619.7$\pm$1.2$(stat)\pm$1.2$(syst)$ MeV/$c^2$.\\

\begin{figure}[htb]
\vspace*{-1mm}
\includegraphics[height=0.30\textheight,width=7.5cm]  {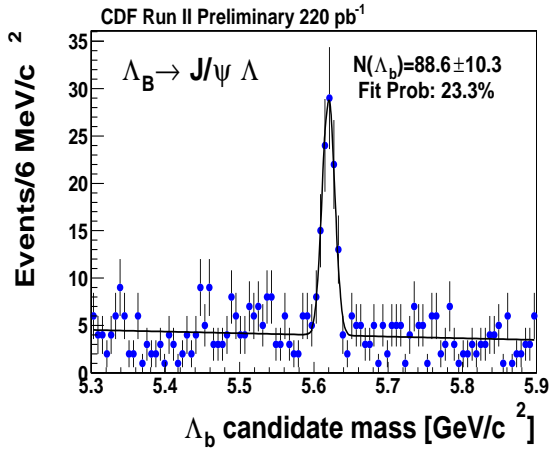}
\vspace*{-1cm}

\caption{The mass spectrum of $\Lambda_b$ candidates (CDF).}
\label{fig:masslb}
\end{figure}

D\O\ reports the first observation of the excited $B$ mesons 
$B_1$ and $B^*_2$ as two separate states in fully reconstructed
decays to $B^{(*)}\pi$. The mass of $B_1$ is measured to be
5724$\pm$4$\pm$7 MeV/c$^2$, and the mass difference $\Delta M$ between
$B^*_2$ and $B_1$ is 23.6$\pm$7.7$\pm$3.9 MeV/c$^2$
(Fig.  \ref{fig:d0_bexc}).

D\O\ observes semileptonic $B$ decays to narrow $D^{**}$ states,
the orbitally excited states  of the $D$ meson
seen as resonances in the $D^{*+}\pi^-$ invariant mass spectrum.
The $D^*$ mesons are reconstructed through the decay sequence 
$D^{*+} \rightarrow D^0\pi^+$, $D^0\rightarrow K^-\pi^+$.
The invariant mass  of oppositely charged $(D^*,\pi)$ pairs
is shown in Fig.  \ref{fig:d0_dstst}.
The mass peak between 2.4 and 2.5 GeV/$c^2$ can be interpreted as two merged 
narrow $D^{**}$ states, $D^0_1(2420)$ and $D^0_2(2460)$.
The combined branching fraction is 
$ {\cal B}(B\rightarrow D^0_1,D^0_2)\cdot {\cal B}(D^0_1,D^0_2\rightarrow D^{*+}\pi^-)=(0.280\pm0.021(stat)\pm0.088(syst)$\%. The systematic error includes the unknown phase between the
two resonances. Work is in progress on extracting the two Breit-Wigner
amplitudes.

\begin{figure}[htb]
\vspace*{-2mm}
\hspace*{-3mm}
\includegraphics[height=0.28\textheight,width=8.3cm]  {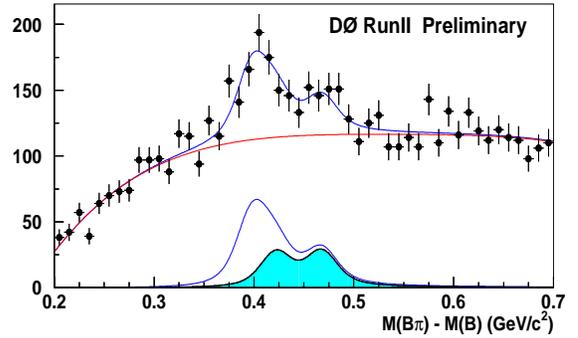}

\vspace*{-1cm}
\caption{Mass difference $\Delta M = M(B\pi)-M(B)$ for exclusive $B$ decays.
The background-subtracted signal is a sum of 
$B^*_1 \rightarrow B^* \pi$, $B^* \rightarrow B \gamma $ (open area)
and $B^*_2 \rightarrow B^*\pi$ $B^*\rightarrow B \gamma$ (lower peak in the shaded area)
and $B^*_2 \rightarrow B \pi$ (upper peak in the shaded area)  
(D\O).}
\label{fig:d0_bexc}
\end{figure}

\begin{figure}[htb]
\includegraphics[height=0.25\textheight,width=7.5cm]  {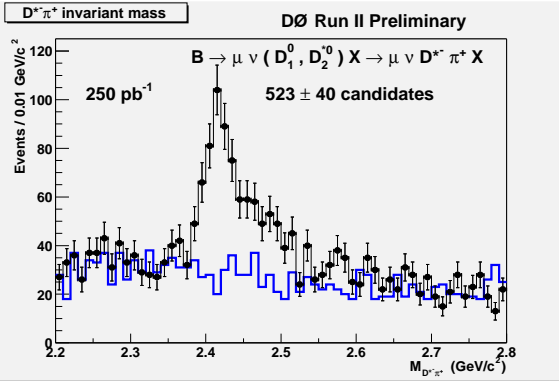}

\vspace*{-1cm}
\caption{The invariant mass distribution of
$(D^*,\pi)$ pairs, opposite sign (points) and same-sign (solid histogram).}
\label{fig:d0_dstst}
\end{figure}

\subsection{Lifetimes}

CDF and D\O\ have measured  lifetimes of $b$ hadrons through the exclusively
reconstructed decays $B^+ \rightarrow J/\psi K^+$, $B^0 \rightarrow J/\psi K^{*0}$,
$B_s \rightarrow J/\psi \phi$, 
and $\Lambda_b \rightarrow J/\psi \Lambda$
(Fig. \ref{fig:d0_lbctau}).
The latest results are:  \\

 $\tau(B^+)$=1.65 $\pm$ 0.08 $^{+0.096}_{-0.123}$  ps ~(D\O\ 2003),

 $\tau(B^+)$=1.662 $\pm$ 0.033  $\pm$ 0.008  ps ~(CDF),

 $\tau(B^0_d)$=1.473  $^{+0.052}_{-0.050}$ $\pm$ 0.023    ps ~(D\O).

 $\tau(B^0_d)$=1.539 $\pm$ 0.051  $\pm$ 0.008  ps ~(CDF),

 $\tau(B^0_s)$=1.444   $^{+0.098}_{-0.090}$ $\pm$ 0.020   ps ~(D\O),

 $\tau(B^0_s)$=1.369 $\pm$ 0.100 $\pm$ $^{+0.008}_{0.010}$  ps ~(CDF),

 $\tau(\Lambda_b)$=1.221 $^{+0.217}_{-0.179}$ $\pm$ 0.043  ps ~(D\O),

 $\tau(\Lambda_b)$=1.25 $\pm$ 0.26 $\pm$ 0.10  ps ~(CDF 2003).\\

The measured lifetimes correspond to the following lifetime ratios:\\

$\tau(B^+)/\tau(B^0_d)$   =  1.080$\pm$0.042     ~(CDF),
 
$\tau(B^0_s)/\tau(B^0_d)$ =  0.890$\pm$0.072    ~(CDF),

$\tau(B^0_s)/\tau(B^0_d)$ = 0.980$ ^{+0.075}_{-0.070}   \pm$0.003    ~(D\O),

$\tau(\Lambda_b)/\tau(B^0_d)$ =  0.874$ ^{+0.169}_{-0.142}   \pm$0.028    ~(D\O).\\

\begin{figure}[htb]
\includegraphics[height=0.3\textheight,width=8.2cm]  {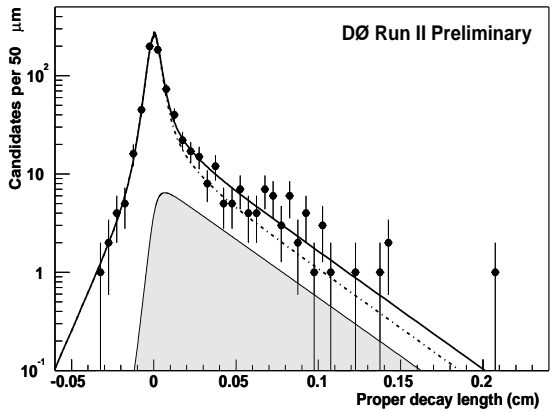}
\vspace*{-1cm}

\caption{ Fit projection on  $c\tau$ for the $\Lambda_b$ candidates.  (D\O)}
\label{fig:d0_lbctau}
\end{figure}

The $B_s$ lifetime measurements listed above are results of
a single-lifetime fit to data, integrated over the decay angles.
Because  of the presence of  final
states  common to \bs\ and its charge conjugate \bsbar,
the two meson states   are expected
to mix in such a way that the two CP  eigenstates may have a relatively
large lifetime difference.
It is possible to
separate the two CP components of \bsdec\ and thus to measure the
lifetime difference by studying the time evolution of the
polarization states of the vector mesons in the final state.
CDF has carried out a combined analysis of $B_s$ lifetimes 
and polarization amplitudes. The results for the lifetimes of the
low mass (CP even) and high mass (CP odd) eigenstates, and the relative 
width difference are:\\

 $\tau_L = 1.05 ^{+0.16}_{-0.13} \pm 0.02$ ~ps,
 
 $\tau_H = 2.07 ^{+0.58}_{-0.46} \pm 0.03$ ~ps,

 $\Delta \Gamma /\overline \Gamma   = 0.65 ^{+0.25}_{-0.33} \pm 0.01$.\\

Figure \ref{fig:cdf_dg} shows  the scan of the likelihood function 
for $\Delta \Gamma /\overline \Gamma$.
Pseudoexperiments tossed with $\Delta \Gamma /\overline \Gamma =0$
yield the betting odds for observing the above results at
1/315. For $\Delta \Gamma /\overline \Gamma = 0.12$ (SM prediction,
which has recently been updated to 0.14$\pm$0.05~\cite{dg_un}) the betting odds are
1/84.

\begin{figure}[htb]
\vspace*{-1mm}
\includegraphics[height=0.3\textheight,width=8.2cm]  {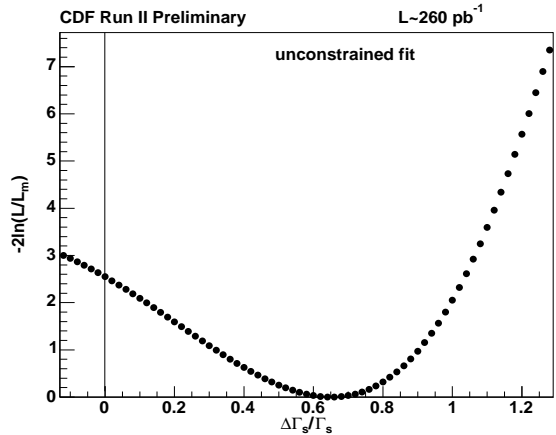}

\vspace*{-1cm}
\caption{Scan of the likelihood function 
for $\Delta \Gamma /\overline \Gamma$ (CDF).
}
\label{fig:cdf_dg}
\end{figure}

D\O\ has used a novel technique to  measure the lifetime ratio
of the charged and neutral $B$ mesons, exploiting the large
semileptonic sample. $B$ hadrons were reconstructed in the channels
$B\rightarrow \mu^+ \nu D^*(2010)^-X$, which are dominated by $B^0$ decays, 
and  $B\rightarrow \mu^+ \nu D^0X$, which are dominated by $B^+$ decays.
The lifetime ratio was
obtained from the variation of the ratio of the number of events in these two
processes at different decay lengths.
The result is \\

$\tau(B^+)/\tau(B^0_d)$   =  1.093$\pm$0.021$\pm$0.022.      ~(D\O)

\subsection{Towards $B_s$ mixing}

Measurement of the $B_s$ oscillation frequency via \bs -\bsbar ~mixing
will provide an important constraint on the CKM matrix. The oscillation
frequency is proportional to the mass difference between the mass eigenstates,
$\Delta m_s$, and is related to the CKM matrix through 
$\Delta m_s \propto |V_{tb}V_{ts}|$. When combined with the
$B_d$ mass difference, $\Delta m_d$ it helps in extraction of $|V_{td}|$,
and thereby the CP violating phase. 

As a benchmark for future $B_s$ oscillation measurement, both groups
study  $B_d$ mixing, gaining an understanding of the different components
of a $B$ mixing analysis (sample composition, flavor tagging, vertexing,
asymmetry fitting). For a sample of partially reconstructed decays
$B\rightarrow D^*(2010)^+\mu^-X$, D\O\ obtains 
$\Delta m_d = 0.506 \pm 0.055 (stat) \pm  0.049 (syst))$ ps$^{-1}$ and
$\Delta m_d = 0.488 \pm 0.066 (stat) \pm  0.044 (syst))$ ps$^{-1}$
when employing  opposite side muon tagging and the same side tagging,
respectively.

The CDF result for semileptonic channels is
$\Delta m_d = 0.536 \pm 0.037 (stat) \pm  0.009 (s.c.) \pm 0.015 (syst)$ ps$^{-1}$.
CDF also reports a result on $B$ oscillations using fully reconstructed
decays:
$\Delta m_d = 0.526 \pm 0.056 (stat) \pm  0.005 (syst))$ ps$^{-1}$.

Reconstructing $B_s$ decays into different final states is another
important
 step in the \bs -\bsbar ~mixing analysis.
Thanks to the  large muon and tracking coverage,   D\O\ is accumulating
a  high statistics sample of semileptonic $B_s$ decays.
D\O\ reconstructs the $B_s \rightarrow D^+_s \mu^- X$ decays, with
$D^+_s \rightarrow \phi \pi^+ $ and
$D^+_s \rightarrow K^* K^+ $,
at a rate of $\approx$ 40(25) events per pb$^{-1}$,  respectively.
Figure \ref{fig:d0_bsdsphipi} shows the mass distribution of the
$D^+_s \rightarrow \phi \pi$ candidates.

\begin{figure}[htb]
\vspace*{-5mm}
\includegraphics[height=0.3\textheight,width=8.0cm]  {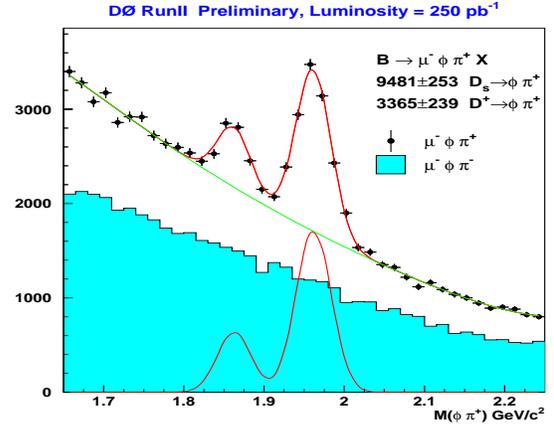}
\vspace*{-1.2cm}
\caption{  $D^+_s \rightarrow \phi \pi^+$  signal. (D\O)}
\label{fig:d0_bsdsphipi}
\end{figure}

\begin{figure}[htb]
\vspace*{-10mm}
\hspace*{-4mm}
\includegraphics[height=0.35\textheight,width=7.9cm]  {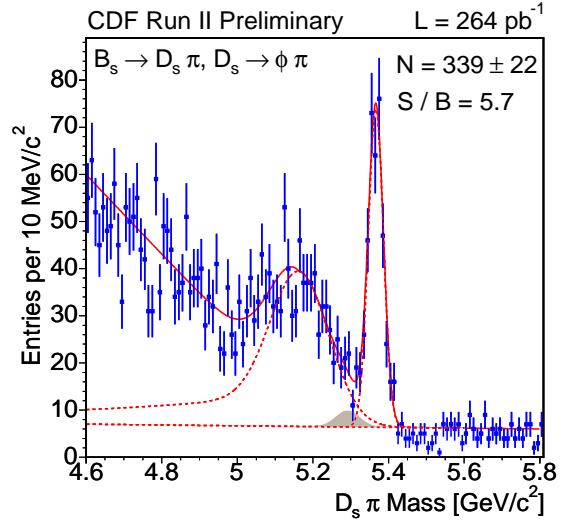}

\vspace*{-1.0cm}
\caption{ $B_s \rightarrow D_s \pi$, $D_s \rightarrow \phi \pi$  signal.  (CDF)}
\label{fig:cdf_bsdsphipi}
\end{figure}

CDF has clean signals for fully hadronic, flavor-specific  $B_s$ decays,
providing the best sensitivity to $B_s$ oscillations at high
$\Delta m_s$. Figure \ref{fig:cdf_bsdsphipi} shows the signal for
the best channel, $B_s \rightarrow D_s \pi$, $D_s \rightarrow \phi \pi$.

\clearpage

\subsection{Rare decays}

The purely leptonic decays $B_{d,s}^0 \rightarrow \mu^+
\mu^-$ are flavor-changing neutral current (FCNC) processes.
In the standard model, these decays are forbidden at the tree level and
proceed at a very low rate through higher-order diagrams.
The latest SM prediction~\cite{sm_ref3}
is ${\cal B}(B^0_s \rightarrow \mu^+ \mu^-)=(3.42\pm 0.54)\times
10^{-9}$, where the error is dominated by non-perturbative uncertainties. The
leptonic branching fraction of the $B_d^0$ decay is suppressed by CKM matrix elements $|V_{td}/V_{ts}|^2$
leading to a predicted SM branching fraction of $(1.00\pm0.14)\times 10^{-10}$.
The best published experimental bound (Fig.~\ref{fig:cdf_bsmumu})
 for the branching fraction
of $B^0_s$ $(B^0_d)$ is presently
${\cal B}(B^0_s \, (B^0_d) \rightarrow \mu^+\mu^-)<7.5\times 10^{-7}\, 
(1.9\times 10^{-7})$ at the 95\% C.L.~\cite{cdfII}.
The decay amplitude of $B^0_{d,s} \rightarrow \mu^+ \mu^-$ can be
significantly enhanced in some extensions of the SM. 

\begin{figure}[htb]
\includegraphics[height=8.3cm,width=7.9cm]  {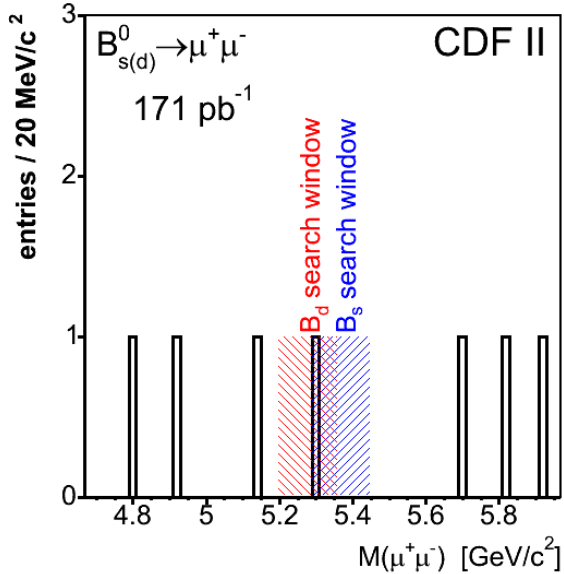}

\vspace*{-1cm}
\caption{Invariant mass for the events passing all requirements. (CDF)}
\label{fig:cdf_bsmumu}
\end{figure}

Assuming no contributions 
from the decay $B^0_d\rightarrow \mu^+\mu^-$ in the signal region,
D\O\  finds the conservative upper limit on the branching fraction 
to be ${\cal B}(B^0_s \rightarrow \mu^+ \mu^-) \leq 4.6\times 10^{-7}$ 
at the 95\% C.L. (Fig.~\ref{fig:d0_bsmumu}).

\begin{figure}[htb]
\includegraphics[height=5.0cm,width=8.0cm]  {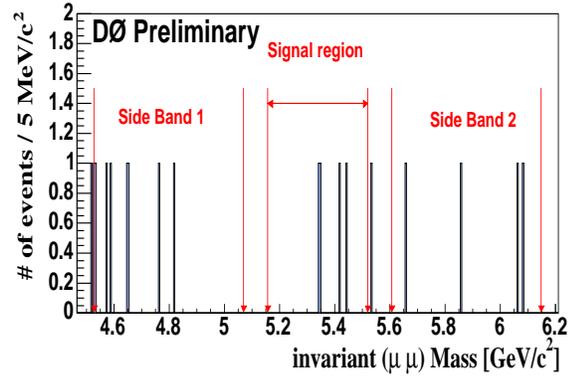}
\vspace*{-1cm}
\caption{Invariant mass for the events  passing all requirements. (D\O)}
\label{fig:d0_bsmumu}
\end{figure}

\end{document}